\documentclass{emulateapj}

\usepackage{apjfonts}

\newcommand{\Md}        {M_{\rm d}}
\newcommand{\Rd}        {R_{\rm d}}
\newcommand{\Mb}        {M_{\rm b}}

\newcommand\degrees{^\circ}

\newcommand{\Msun}{\mbox{$\rm M_{\odot}$}}
\newcommand{\amp}[1]{\mbox{$A_{#1}$}}

\newcommand{\omg}[1]{\mbox{$\Omega_{#1}$}}
\newcommand{\omtw}{\mbox{$\omg{TW}$}}

\newcommand{\pin}[1]{\mbox{${\cal X}_{#1}$}}
\newcommand{\kin}[1]{\mbox{${\cal V}_{#1}$}}
\newcommand{\sigpin}{\mbox{${\sigma_{\cal X}}$}}
\newcommand{\sigkin}{\mbox{${\sigma_{\cal V}}$}}
\newcommand{\sigom}{\mbox{${\sigma_\Omega}$}}
\newcommand{\panuc}{\mbox{$\psi_{\rm nuc}$}}

\newcommand{\ie}{{\it i.e.}}

\shorttitle{Double-Barred Galaxies in $N$-body Simulations}
  
\shortauthors{Shen \& Debattista}

\begin{document}
\title{Observable Properties of Double-Barred Galaxies in $N$-Body Simulations}

\author{Juntai Shen\altaffilmark{1}} 
\affil{McDonald Observatory, The University of
Texas at Austin, 1 University Station, C1402, Austin, TX 78712}
\altaffiltext{1}{Harlan J. Smith Fellow}
\email{shen@astro.as.utexas.edu}

\and

\author{Victor P. Debattista\altaffilmark{2}} 
\affil{Astronomy Department, University of Washington, Seattle, WA 98195}
\altaffiltext{2}{Brooks Prize Fellow; current address: Centre For
Astrophysics, University of Central Lancashire, Preston, UK PR1 2HE}
\email{debattis@astro.washington.edu}

\slugcomment{Draft Version of  \textit{\today}}

\begin{abstract}
Although at least one quarter of early-type barred galaxies host
secondary stellar bars embedded in their large-scale primary
counterparts, the dynamics of such double barred galaxies are still
not well understood.  Recently we reported success at simulating such
systems in a repeatable way in collisionless systems.  In order to
further our understanding of double-barred galaxies, here we
characterize the density and kinematics of the $N$-body simulations of
these galaxies.  This will facilitate comparison with observations and
lead to a better understanding of the observed double-barred
galaxies. We find the shape and size of our simulated secondary bars
are quite reasonable compared to the observed ones. We demonstrate
that an authentic decoupled secondary bar may produce only a weak twist of
the kinematic minor axis in the stellar velocity field, due to the
relatively large random motion of stars in the central region. We also
find that the edge-on nuclear bars are probably not related to boxy
peanut-shaped bulges which are most likely to be edge-on primary
large-scale bars. Finally we demonstrate that the non-rigid rotation
of the secondary bar causes its pattern speed not to be derived with
great accuracy using the Tremaine-Weinberg method.  We also compare
with observations of NGC 2950, a prototypical double-barred early-type
galaxy, which suggest that the nuclear bar may be rotating in the
opposite sense as the primary.
\end{abstract}


\section{Introduction} 
\label{sec:intro}

Double barred (S2B) galaxies were first described over thirty years
ago \citep{devauc_75}.  The {\it Hubble Space Telescope} has revealed
secondary bars at the center of at least one quarter of early-type
optically-barred galaxies \citep{erw_spa_02}.  Dynamically
decoupled\footnote{In this context, by decoupled we mean only that
$\omg{s} \neq \omg{p}$, where \omg{s} (\omg{p}) is the pattern speed
of the secondary (primary) bar.}  secondary bars in S2B galaxies have
been hypothesized to be a mechanism for driving gas past the inner
Lindblad resonance (ILR) of primary bars, to feed the supermassive
black holes that power active galactic nuclei \citep{shl_etal_89}.

The dynamics of secondary bars are still not well understood. The
random apparent relative orientations of primary and secondary bars in
nearly face-on galaxies points to dynamical decoupling
\citep{but_cro_93, fri_mar_93}.  But images alone cannot reveal much
about how the two bars rotate through each other.  Kinematic evidence
of decoupling, using either gas or stars, is harder to obtain
\citep{pet_wil_02, sch_etal_2002,moi_etal_04}.  Indirect evidence for
decoupling was claimed by \citet{ems_etal_01} based on rotation
velocity peaks inside the secondary bars.  Conclusive direct kinematic
evidence for a decoupled secondary bar was obtained for NGC 2950 by
\citet[][hereafter CDA03]{cor_deb_agu_03} who showed, using the method
of \citet{tre_wei_84}, that the primary and secondary bars cannot be
rotating at the same pattern speed.

\begin{figure*}[!ht]
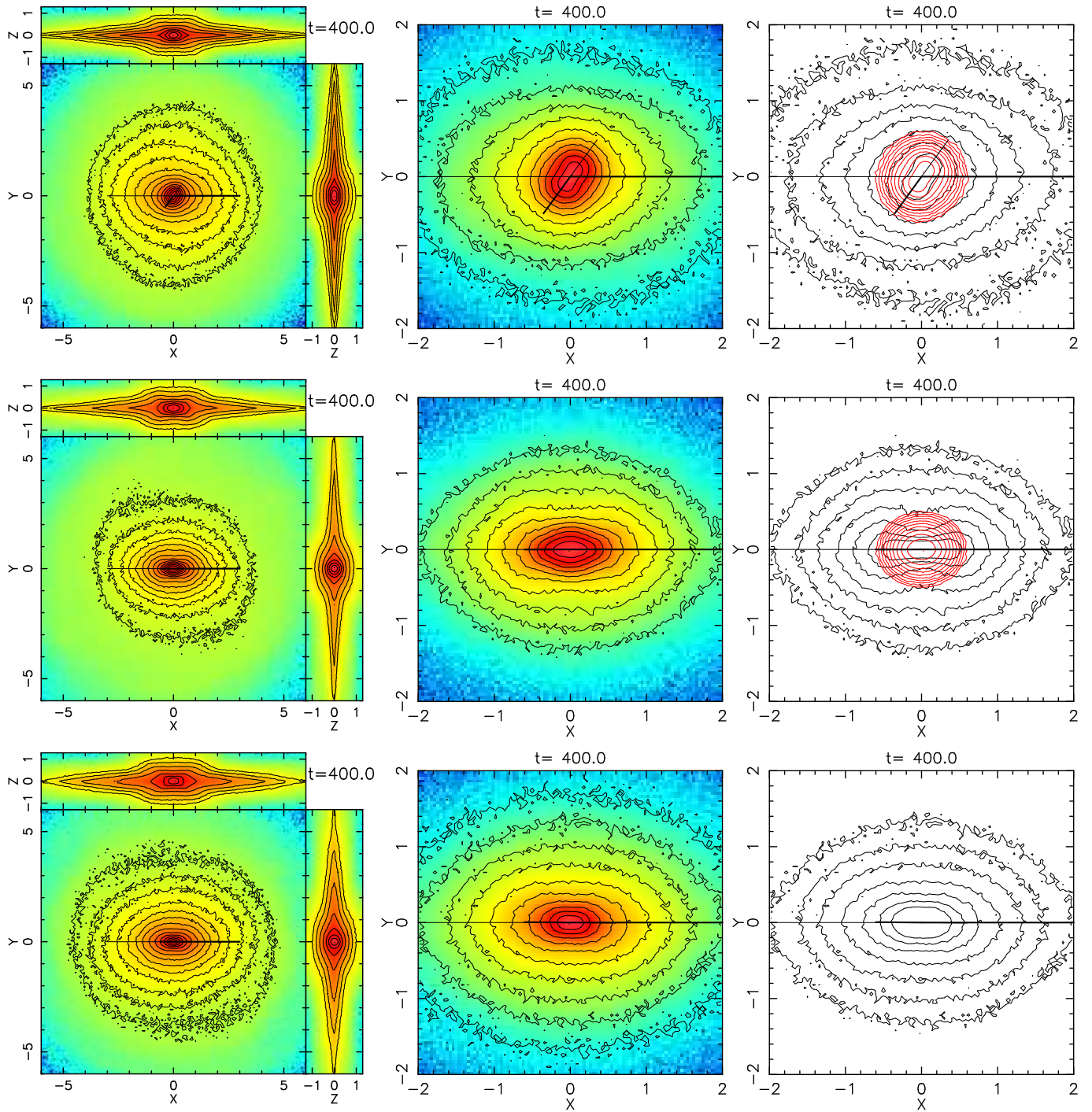

\centerline{
\includegraphics[angle=-90.,width=0.345\hsize]{fig1aa.ps}
\includegraphics[angle=-90.,width=0.325\hsize]{fig1ab.ps}
\includegraphics[angle=-90.,width=0.325\hsize]{fig1ac.ps}}
\vspace{0.01\hsize}
\centerline{
\includegraphics[angle=-90.,width=0.345\hsize]{fig1ba.ps}
\includegraphics[angle=-90.,width=0.325\hsize]{fig1bb.ps}
\includegraphics[angle=-90.,width=0.325\hsize]{fig1bc.ps}}
\vspace{0.01\hsize}
\centerline{
\includegraphics[angle=-90.,width=0.345\hsize]{fig1ca.ps}
\includegraphics[angle=-90.,width=0.325\hsize]{fig1cb.ps}
\includegraphics[angle=-90.,width=0.325\hsize]{fig1cc.ps}}

\caption{3-D surface density of 3 runs. In each row, from left to
 right, panels are for the large-scale total stellar distribution,
 central zoom-in of total stellar distribution, separate disk (dark)
 and bulge (red) surface density contours, respectively.  (a). {\it
 top row}: the canonical run with a secondary bar formed (Run D);
 (b). {\it middle row}: the run with an unrotating bulge, which
 formed only a single bar (run S); (c). {\it bottom row}: the run
 without a bulge (run NB). The contours in $x$-$y$ plane are separated
 by half dex, while in vertical projections they are spaced more
 sparsely to avoid contour over-crowding.}
\label{fig:3dshape}
\end{figure*}

Simulations offer the best way to understand double barred
systems. However, the decoupled nuclear bars that formed in early
simulations did not last long.  For example, the most long-lived
nuclear bar in \citet{fri_mar_93} lasted for less than two turns of
the primary bar, corresponding to about 0.4 Gyr, which is far too
short to explain the observed abundance of nested bars.  Furthermore,
their models usually require substantial amounts of gas to form and
maintain these nuclear bars. \citet{hel_etal_07, hel_etal_07_2}
reported that nested bars form in a quasi-cosmological setting, but the
amplitudes of the bars also seem to weaken rapidly after most of gas
has formed stars \citep[][Figure~2]{hel_etal_07}. \citet{pet_wil_04}
found that 4 out of 10 double-barred galaxies contain very little
molecular gas in the nuclear region. These clues suggest that large
amounts of molecular gas may not be necessary to maintain central
nuclear bars. \citet{rau_etal_02} reported that a secondary bar forms
in a collisionless $N$-body simulation, although their secondary bar
had a ``vaguely spiral shape.''

On the side of orbital studies, \citet{mac_spa_97,mac_spa_00}
discovered a family of loop orbits that may form building blocks of
long-lived nuclear stellar bars (also \citealt{mac_ath_07}). Their
studies are very important for the understanding of double barred
galaxies, but their models are not fully self-consistent, since nested
bars in general cannot rotate rigidly through each other
\citep{lou_ger_88}. So fully self-consistent $N$-body simulations are
still needed to check if their main results still hold when the
non-rigid nature of the bars is taken into account.

Recently \citet[hereafter DS07]{deb_she_07} demonstrated that
long-lived secondary bars can form in purely collisionless $N$-body
simulations, when a rotating pseudobulge is introduced in their
model. The nuclear bars in their work are distinctly bars, and do not
have a spiral shape. They showed that the behavior of their models
were in good agreement with the loop orbit predictions of
\citet{mac_spa_00}.

In this report we analyze the photometrical and kinematical properties
of high resolution models in detail. Our theoretical results here can
also be compared to the observed 2-D kinematics of some double-barred
galaxies, to achieve a better understanding of the dynamics of the
secondary bars.


\section{Models}
\label{sec:models}

The simulations presented in this paper are all collisionless. The
model setup is very similar to that of DS07. As in
DS07, the formation of the secondary bar is induced by a
rotating pseudobulge.  We focus on three simulations: run D which
formed a long-lasting double-barred system due to an initially
rotating bulge, run S in which only a single bar formed with an
initially unrotating bulge component, and run NB where there is no
bulge component initially. Our high-resolution simulations consist of
live disk and bulge components in a rigid halo potential.  We restrict
ourselves to rigid halos to allow higher mass resolution in the
nuclear regions, to study the complicated co-evolution of the two bars
without the additional evolution introduced by the halo.  The rigid
halos used in this study are all logarithmic potentials $\Phi(r) =
\frac{1}{2}V_{\rm h}^2~ \ln(r^2 + r_{\rm h}^2)$. We set $V_{\rm h} =
0.6$ and $r_{\rm h} = 15$ in all runs. We employed about four times
more particles than the runs published in DS07 to better
analyze the photometric and kinematic properties; Run D and S have
$4.8 \times 10^6$ equal mass particles, with $4 \times 10^6$ in the
disk and the rest in the bulge. Run NB has $4 \times 10^6$ in the disk
only since there is no bulge.

The initial disks in our simulations all have exponential surface
densities with scale-length $\Rd$, mass $\Md$ and Toomre-$Q\simeq
2$. The bulge was generated using the method of \citet{pre_tom_70} as
described in \citet{deb_sel_00}, where a distribution function is
integrated iteratively in the global potential, until convergence.  In
both run D and run S the bulge has mass $\Mb=0.2\Md$ and we used an
isotropic King model distribution function.  The bulge truncation
radius is $0.9\Rd$ in both run D and run S.  The bulge set up this way
is non-rotating.  We introduce bulge rotation in run D by simply
reversing the velocities of bulge particles with negative angular
momenta, which is still a valid solution of the collisionless
Boltzmann equation \citep{lynden_62}.  The bulges in run D and run S
are flattened by the disk potential initially, and remain so at later
times. The initial kinematic ratio $V_p/\bar\sigma$ in run D is
slightly above the line for oblate isotropic rotators \citep{binney_78}.

We use $\Rd$ and $\Md$ as the units of length and mass, respectively,
and the time unit is $(\Rd^3/G\Md)^{1/2}$.  If we scale these units to
the physical values $\Md = 2.3 \times 10^{10} \Msun$ and $\Rd = 2.5$
kpc, then a unit of time is $12.3$ Myr.  We use a force resolution
(softening) of $0.01$, which scaled to the above physical units
corresponds to 25 pc.  These simulations were evolved with a 3-D
cylindrical polar grid code \citep{sel_val_97}.  This code expands the
potential in a Fourier series in the cylindrical polar angle $\phi$;
we truncated the expansion at $m=8$. Forces in the radial direction
are solved for by direct convolution with the Greens function while
the vertical forces are obtained by fast Fourier transform.
We used grids measuring $N_R\times N_\phi \times N_z = 58 \times 64
\times 375$.  The vertical spacing of the grid planes was $\delta z =
0.01 R_{\rm d}$.  Time integration used a leapfrog integrator with a
fixed time step $\delta t = 0.04$.


\section{Photometry}
\label{sec:photometry}

\subsection{Shape of the secondary bar}

Figure~\ref{fig:3dshape} shows the surface density contours and images
of the double-barred run D, run S, and run NB which does not have an
initial bulge component. From Figure~\ref{fig:3dshape} we see that for
run D the secondary bar shows up in both the disk and bulge
components. All three large scale bars appear qualitatively similar to
each other in both the face-on and edge-on views.

When viewed side-on, the large scale bar appears to be
boxy/peanut-shaped regardless of whether or not an initial live bulge
is included, or if a secondary bar is present. The formation of a
boxy/peanut-shaped bulge from disk has been studied extensively with
$N$-body simulations
\citep[e.g.][]{com_etal_90,rah_etal_91,bur_ath_05,deb_etal_05,mar_etal_06,deb_etal_06}.
Figure~\ref{fig:3dshape} does not show any obvious influence of the
secondary bar on the overall boxy/peanut-shaped side-on appearance of
a large-scale bar. So it is quite unlikely that most boxy-shaped
bulges are edge-on nuclear bars, as speculated by \citet{kor_ken_04}
as a possibility of explaining boxy bulges. This is hardly surprising
as the small size of the secondary bar makes its side-on signatures,
if any, easily masked by the primary. A caveat may be that the
boxiness in Figure~\ref{fig:3dshape} does not cover a range as wide as
that in \citet{deb_etal_06}. Also note that the boxy part is smaller
than the primary bar (regardless of double-barred or single-barred) as
a whole \citep[see][]{she_sel_04,kor_ken_04,mar_etal_06, deb_etal_06,
ath_bea_07}.

Figure~\ref{fig:n2950mod} shows the projected system (at $t=405$ when
the two bars are nearly perpendicular) with an ordinary orientation:
the system is inclined at $i = 45\degrees$ with the line of nodes
(LON) of $\panuc =45\degrees$ relative to the secondary bar major
axis.  The surface density image and contours resemble many observed
double-barred systems, such as NGC 2950, even though we did not
deliberately set out to match it.

\begin{figure}
\includegraphics[angle=-90.,width=1.\hsize]{fig2.ps}
\caption{Run D at $t=405$ projected to $i = 45\degrees$ and $\panuc =
45\degrees$ with all particles shown.  The model bears a passing
resemblance to NGC 2950 \citep{erw_spa_02}.
\label{fig:n2950mod}}
\end{figure}

\begin{figure}[!ht]
\centerline{
\includegraphics[angle=0.,width=\hsize]{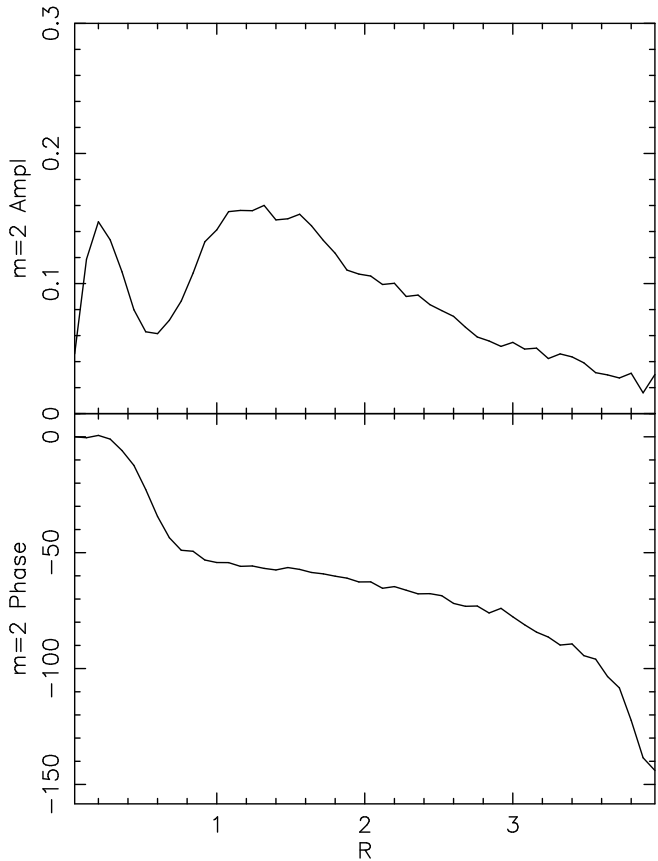}}
\caption{The radial variations of the $m=$2 Fourier amplitude and phase of all particles for Run D at $t=400$.}
\label{fig:m_2}
\end{figure}

\begin{figure}[!ht]
\centerline{
\includegraphics[angle=0.,width=\hsize]{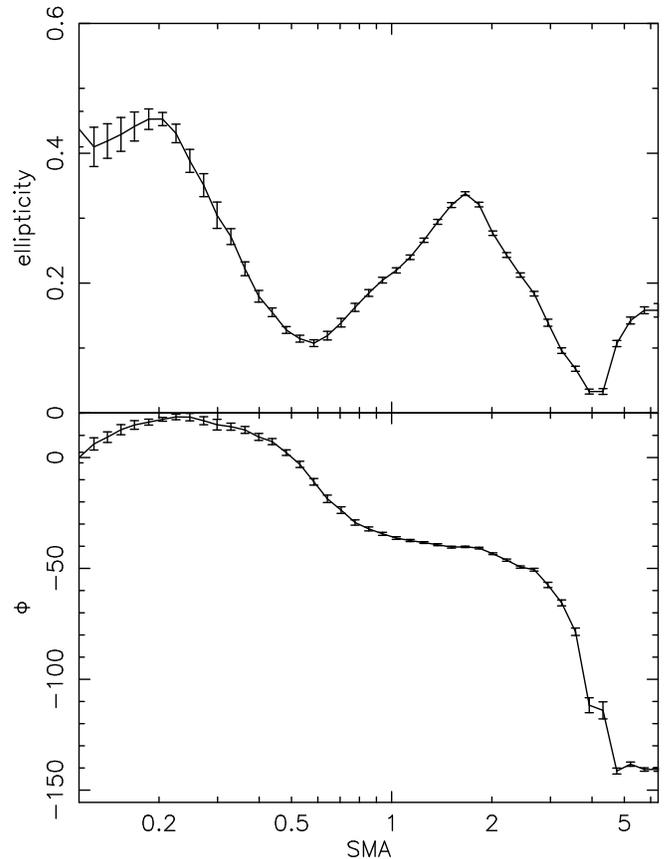}}

\caption{Ellipticity and position angle as a function of semi-major axis of IRAF-fitted ellipses for Run D at $t=400$.}
\label{fig:iraf}
\end{figure}

\subsection{Size relation of the two bars}
\label{sec:barlength}
Figure~\ref{fig:m_2} shows radial variations of $m=2$ Fourier
amplitude and phase for run D at $t=400$. Figure~\ref{fig:iraf} shows
the ellipticity and position angle (PA) profiles of ellipses fitted
with IRAF for the same data as in Figure~\ref{fig:m_2} (we use log
scale for radius to be consistent with what observers usually
adopt). There are four popular methods for determining the semi-major
axis $a_B$ of a bar.  As summarized by \citet{one_dub_03} and
\citet{erwin_05}. For convenience, we denote the primary bar as B1 and
the secondary bar as B2.

(1) the bar end is measured by extrapolating half-way down the slope
    on the $m=2$ amplitude plot (Fig~\ref{fig:m_2}a). We find $a_{B1}
    \sim 2.3$, $a_{B2} \sim 0.4$, the B2/B1 bar length ratio is about
    $\sim 0.17$.

(2) the bar end is measured when $m=2$ phase deviates from a constant
    by $10^\circ$ (Fig~\ref{fig:m_2}b). We find $a_{B1} \sim 2.1$,
    $a_{B2} \sim 0.4$, the B2/B1 bar length ratio is about $\sim
    0.19$.

(3) the bar end is measured at the peak of the fitted ellipticity
    profiles \citep[e.g.][]{mar_jog_07,men_etal_07}, which is shown in
    Fig~\ref{fig:iraf}a. We find $a_{B1} \sim 1.7$, $a_{B2} \sim 0.2$,
    the B2/B1 bar length ratio is about $\sim 0.12$.

(4) the bar end is measured when the PA of fitted ellipses deviates
    from a constant by $10^\circ$. (Fig~\ref{fig:iraf}b). We find
    $a_{B1} \sim 2.3$, $a_{B2} \sim 0.4$, the B2/B1 bar length ratio
    is about $\sim 0.17$.

Method 1, 2 and 4 yield consistent values of the bar lengths and
length ratios. We found that method 3 tends to give a lower value of
bar lengths than the other three methods, as shown in
\citet{one_dub_03}. Although these methods have some uncertainties in
measuring the bar lengths, the length ratio of the two bars is in the
range of 0.12 to 0.19 (in particular method 1, 2, and 4 give a
consistent narrow range of 0.17 to 0.19). This result is in good
agreement with the typical observed length ratio of local S2B systems
(median ratio $\sim$ 0.12, see \citealt{erw_spa_02, erwin_04,
lis_etal_06}).  Note that we expect that the length of the secondary
cannot be too large, otherwise the gravitational torque from the
primary bar will inevitably twist the secondary into alignment if they
rotate at different pattern speeds.

\subsection{Face-on surface density profiles} 

\begin{figure*}[!ht]
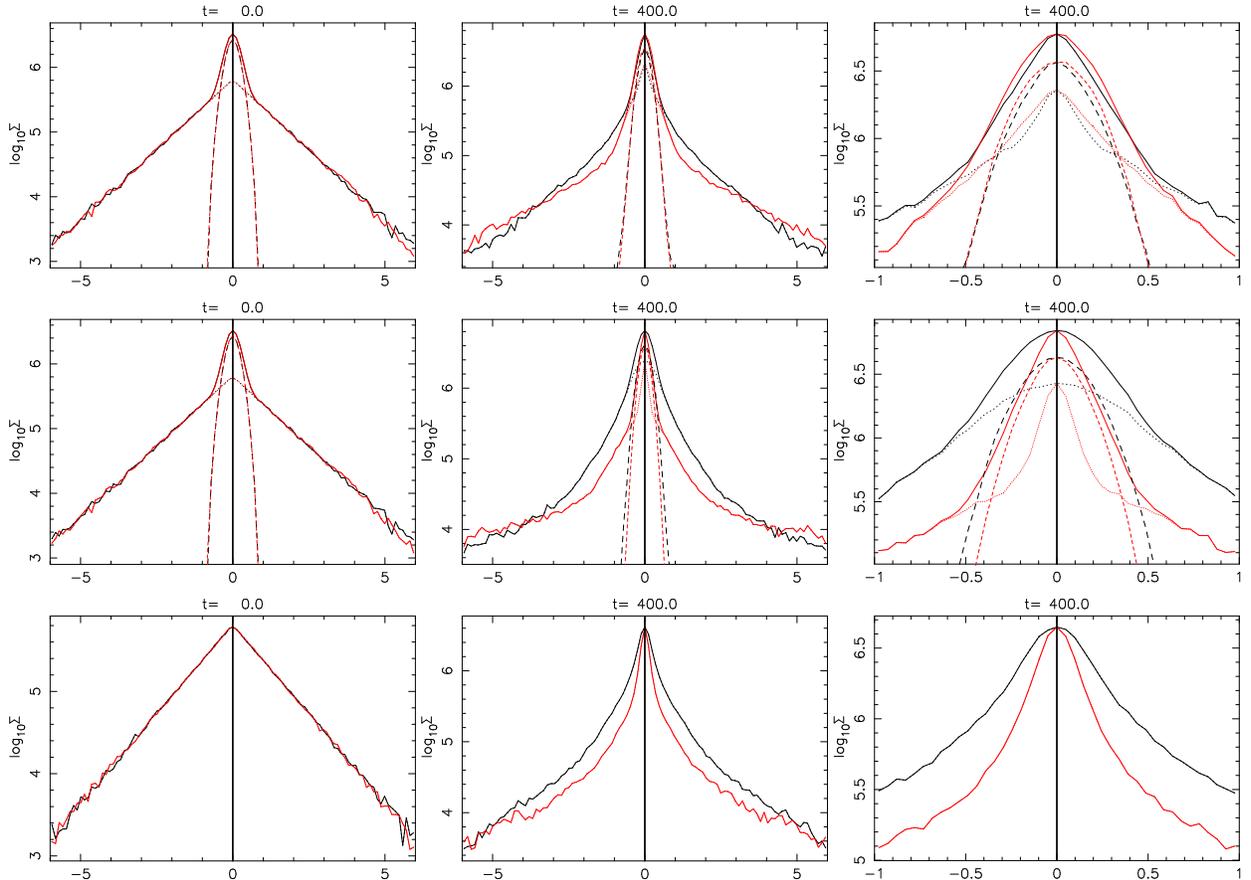

\centerline{
\includegraphics[angle=-90.,width=0.3\hsize]{fig5aa.ps}
\includegraphics[angle=-90.,width=0.3\hsize]{fig5ab.ps}
\includegraphics[angle=-90.,width=0.3\hsize]{fig5ac.ps}}
\vspace{0.01\hsize}
\centerline{
\includegraphics[angle=-90.,width=0.3\hsize]{fig5ba.ps}
\includegraphics[angle=-90.,width=0.3\hsize]{fig5bb.ps}
\includegraphics[angle=-90.,width=0.3\hsize]{fig5bc.ps}}
\vspace{0.01\hsize}
\centerline{
\includegraphics[angle=-90.,width=0.3\hsize]{fig5ca.ps}
\includegraphics[angle=-90.,width=0.3\hsize]{fig5cb.ps}
\includegraphics[angle=-90.,width=0.3\hsize]{fig5cc.ps}}
\caption{Face-on surface density profiles along major and minor axis
of the primary bar for run D (top row), run S (middle row) and run NB
(bottom row). For each row, from left to right: the surface density
profile at $t=0$, $t=400$, and the close-up view of the inner region
at $t=400$, respectively. In all figures, solid lines are surface
density for all particles, dashed lines are for bulge particles only,
dotted are for disk particles only. The black and red curves are along
the major and minor axis of the primary bar, respectively.}
\label{fig:sd}
\end{figure*}

Figure~\ref{fig:sd} shows the face-on surface density profiles along
the major and minor axis of the primary bar in run D, S, and
NB. Compared with the initial profile of each run, there is a
significant increase in central density following the formation of the
large scale bar which redistributes the disk particles. For run D, the
central density profile along the primary bar major axis is no longer
higher than that along the minor axis, due to the secondary bar
orienting to a different direction from the primary bar. This minor
axis over-density is, of course, even more pronounced when the two
bars are perpendicular. This can be an important signature of
confirming small secondary bars photometrically, especially when the
central region is not well resolved. Variations in M/L are unlikely to
mask this minor/major axis difference as density profiles are for
roughly the same radial range. On the other hand, long wavelength
photometry is preferred to minimize the effects of dust.  We do not
find other significant differences in the face-on surface density
profiles between run D and other runs without a secondary bar.

\section{Kinematics}
\label{sec:kinematics}

\begin{figure}[!t]
\centerline{
\includegraphics[angle=-90.,width=\hsize]{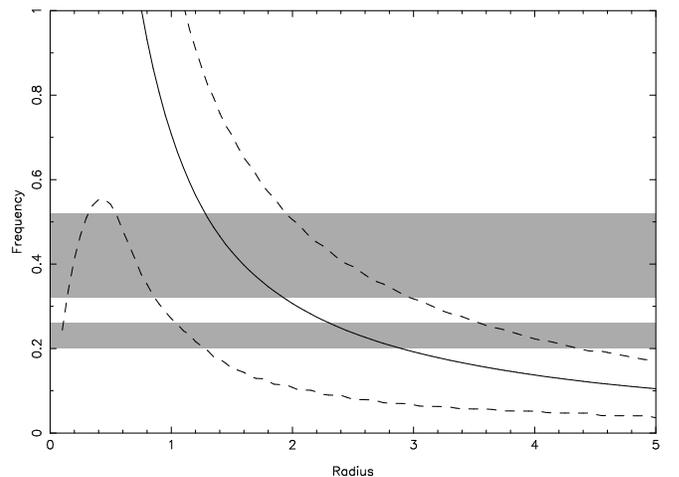}}

\caption{Frequencies as a function of radius at around $t=400$ for run
D, calculated based on the azimuthally averaged gravitational
attraction. The full-drawn line shows the curve of the circular
angular frequency $\Omega$ and the dashed curves mark
$\Omega\pm\kappa/2$, where $\kappa$ is the epicyclic frequency. The
two shaded bands show the oscillational ranges of the bar
pattern speeds (the upper band is for the secondary bar and the lower
one is for the primary). }
\label{fig:spct}
\end{figure}

Figure~\ref{fig:spct} shows the behavior of the azimuthally averaged
$\Omega$, $\Omega\pm\kappa/2$, and the location of the Lindblad
resonances of the bars at around $t=400$ for run D. As shown in
DS07, the pattern speeds of the bars, especially that of
the secondary, vary as they rotate through each other: the secondary
bar rotates slower than average when the two bars are perpendicular,
and faster when the bars are parallel. The patten speed bands shown in
Figure~\ref{fig:spct} reflect such variations. Clearly the pattern
speed of the secondary bar oscillates much more than that of the
primary. The primary bar extends roughly to its CR radius ($\sim2.5$),
consistent with the general expectation and is therefore considered a
fast bar \citep[e.g. CDA03,][]{deb_wil_04}. The secondary bar rotates
faster than the primary bar. However, the secondary bar is much
shorter than its shortest $R_{\rm CR}$. In addition, even if the
variation of the pattern speed is taken into account, the $R_{\rm CR}$
of the secondary is not very close to the $R_{\rm ILR}$ of the
primary, if we use the same naive definition of $R_{\rm ILR}$ as in
\citet{pfe_nor_90}\footnote{A cautionary note is that the $R_{\rm
ILR}$ read naively from Figure~\ref{fig:spct} serves just as a visual
guide, because the $R_{\rm ILR}$ determined this way is reliable
only for weak bars, which is questionable for our strong bars
\citep[e.g.,][]{vanalb_san_82}.}. This is inconsistent with the
speculated CR-ILR coupling requirement for making secondary bars
\citep[e.g.][]{pfe_nor_90,fri_mar_93}.

Since the long-lasting secondary bars which form in our simulations
generally do not extend to their corotation radii, the kinematics
of secondary bars differ from those of primary bars in at least this
important detail.  We here explore the kinematic observables of S2Bs
in our simulations in more detail.

\subsection{Line-of-sight velocity distribution}

We analyzed the line-of-sight velocity distribution (LOSVD) by
measuring the mean velocity $\overline{v}$ and velocity dispersion
$\sigma$. Departures from a Gaussian distribution are parameterized by
Gauss-Hermite moments \citep{gerhar_93, van_fra_93,ben_etal_94}. The
second order term in such an expansion is related to the
dispersion. Following \citet{gerhar_93} the third-order term $h_3$ and
fourth-order term $h_4$ are defined as

$$h_n=\frac{\sqrt{4\pi}}{\Sigma} \int l(w)H_n(w) \exp^{(-1/2)w^2}dw$$
where $w=(v-\overline{v})/\sigma$, $n=3$ or $4$,
$H_3(w)=[1/(96\pi)^{1/2}](8w^3-12w)$, and
$H_4(w)=[1/(768\pi)^{1/2}](16w^4-48w^2+12)$. 
For a particle model, the integral becomes a sum and $\Sigma$ is
replaced by $N_p$, the number of particles in a bin. $h_3$ measures
deviations that are asymmetric about the mean, while $h_4$ measures
the lowest order symmetric deviations from Gaussian (negative for a
``flat-top'' distribution, and positive for a more peaked one).

\begin{figure*}[!t]
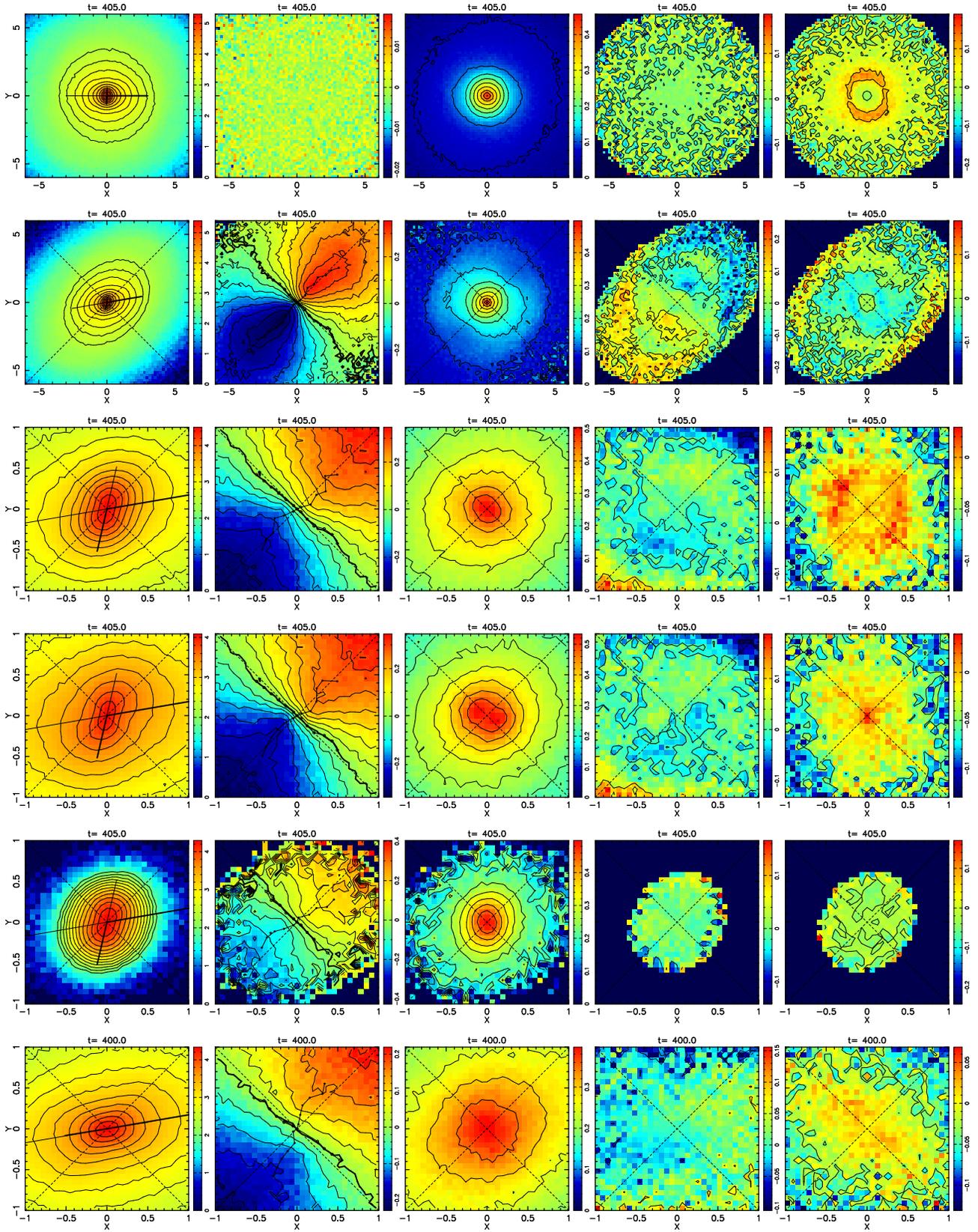

\centerline{
\includegraphics[angle=-90.,width=0.95\hsize]{fig7a.ps}}
\vspace{0.01\hsize}
\centerline{
\includegraphics[angle=-90.,width=0.95\hsize]{fig7b.ps}}
\vspace{0.01\hsize}
\centerline{
\includegraphics[angle=-90.,width=0.95\hsize]{fig7c.ps}}
\vspace{0.01\hsize}
\centerline{
\includegraphics[angle=-90.,width=0.95\hsize]{fig7d.ps}}
\vspace{0.01\hsize}
\centerline{
\includegraphics[angle=-90.,width=0.95\hsize]{fig7e.ps}}
\vspace{0.01\hsize}
\centerline{
\includegraphics[angle=-90.,width=0.95\hsize]{fig7f.ps}}
\vspace{0.01\hsize}

\caption{Photometrical and kinematic maps of run D and run NB. For
each row from left to right are the projected surface density, mean
velocities (``spider diagrams''), velocity dispersion, $h_3$, and
$h_4$ maps. (a): Row 1, run D at $t=405$ face-on. (b): Row 2, run D at
$t=405$ inclined at $45^\circ$ with the LON of $45^\circ$ relative to
the primary bar major axis; (c): Row 3, close-up views of run D at
$t=405$ projected the same way as (b); (d): Row 4, as in (c) but
include disk particles only; (e): Row 5, as in (c) but include bulge
particles only; (f): Row 6, run NB at $t=400$ inclined at $45^\circ$
with the LON of $45^\circ$ relative to the (single) bar major
axis. The corresponding fields for run S are similar to run NB so not
shown for brevity; The short and long straight line segments labels
the direction of the secondary and primary bar, respectively (note the
length of the line segment does not represent the bar length). For the
projected plot, one of the dashed lines represents PA of the line of
nodes ($45^\circ$), while the other dashed is the anti-PA
($135^\circ$). In the mean velocity map, the line with the connected
dots shows the rough position of kinematic major axis, while the heavy
solid curve is the zero velocity curve (kinematic minor axis). The
$h_3$ and $h_4$ analyses are not preformed to bins with less than 100
particles. }
\label{fig:kinemap}
\end{figure*}

Figure~\ref{fig:kinemap}(a-e) show the images and the LOS stellar
kinematics of run D at $t=405$ when the two bars are almost
perpendicular to each other. For comparison purpose
Figure~\ref{fig:kinemap}(f) is for run NB (run S is very similar to
run NB, so it is not shown for brevity). As in
Figure~\ref{fig:n2950mod}, we project the system to $i=45^\circ$ with
the LON of $45^\circ$ relative to the secondary bar major axis.

The most striking feature in Figure~\ref{fig:kinemap} is that the
twist of the kinematic minor axis (i.e. $v_{\rm los}=0$) in the
secondary bar region is weak (see the mean velocity maps). The
kinematic minor axis is almost perpendicular to the inclination axis,
although there is a small but noticeable twisted pinch near the
kinematic minor axis in the nuclear region. The weak central twist is
mainly due to the relatively large velocity dispersion, especially in
the central region (likewise at $t=20$ when only the small nuclear bar
exists, the stellar twist is slightly stronger than at $t=405$, but
still quite small compared to the expected twist in gaseous
kinematics).  On the other hand, the twist of the kinematic {\it
major} axis is more prominent in the central region.
\citet{moi_etal_04} found the stellar kinematic minor axis hardly
twists from the PA of the disk in their sample with the most reliable
kinematics, leading them to question whether nuclear photometric
isophotal twists represent {\it bona fide} dynamically decoupled
secondary bars. We demonstrate that an authentic decoupled secondary
bar may indeed produce a very weak twist of the kinematic minor axis
in the stellar velocity field. So a central stellar velocity map
without a strong twist as in \citet{moi_etal_04} does not necessarily
exclude the existence of a decoupled nuclear bar.

As a comparison, Figure~\ref{fig:kinemap}(f) shows that the
kinematical minor axis twist is just slightly stronger for the
single-barred run NB.  Of course gas kinematics may show much more
twisted features than the stellar data
\citep[e.g.][]{moi_etal_04,ems_etal_06}. However, the gas in the
nuclear region is more prone to non-gravitational forces like shocks,
AGN jets and outflows, so may not directly probe the underlying
gravitational potential.

It is also worth noting that the $\sigma$ symmetry axis does not align
with the secondary or the primary bar (also true for the single bar
run NB), which is consistent with what \citealt{moi_etal_04} found.
We do not find a clear signature in the $h_3$ map associated with
secondary and primary bar. The ring in the $h_4$ map of
Figure~\ref{fig:kinemap}(c) is not always present for different
projections, so it cannot be used to detect secondary bars.

\begin{figure*}[!t]
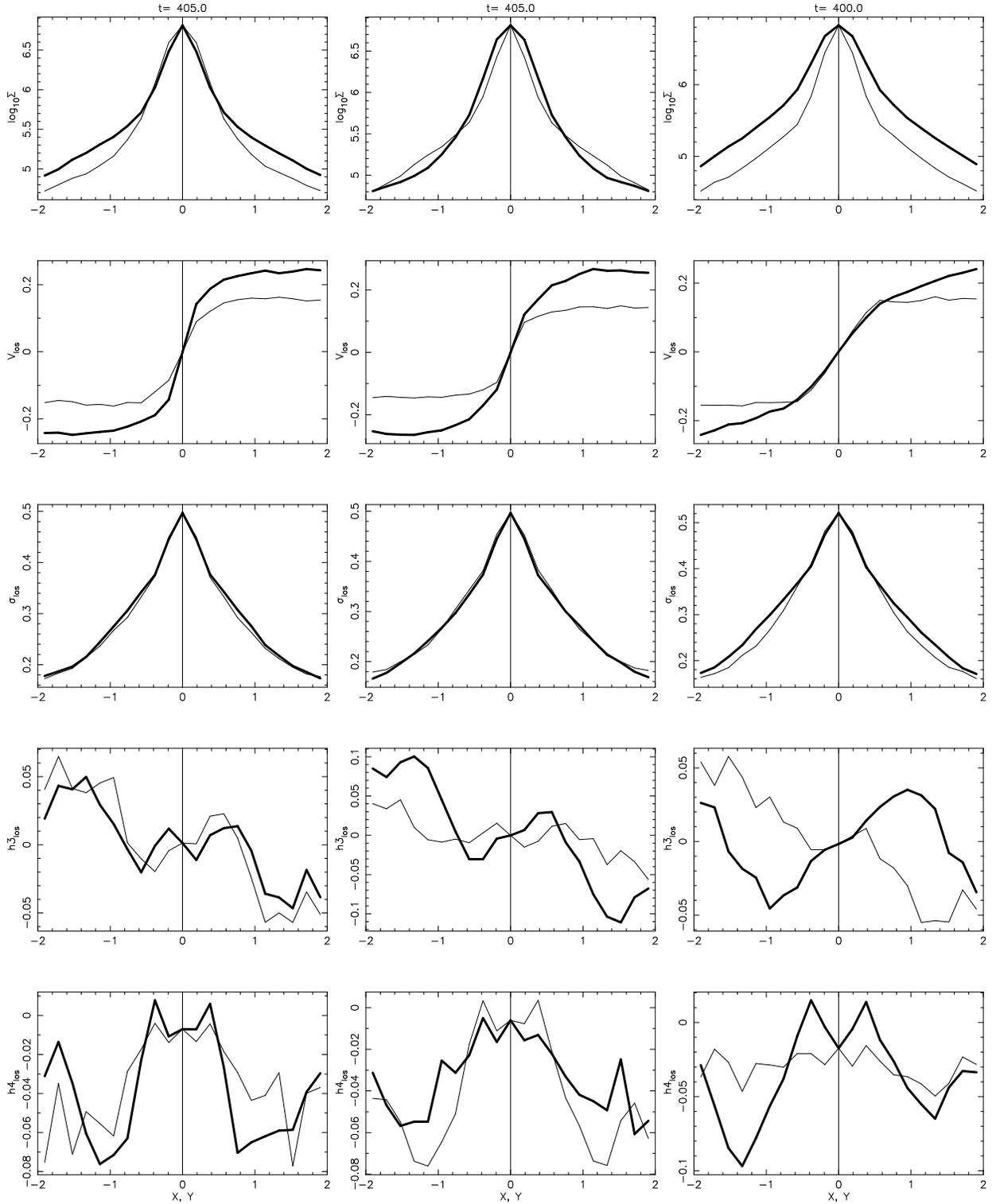

\centerline{
\includegraphics[angle=-90.,width=0.3\hsize]{fig8a.ps}
\includegraphics[angle=-90.,width=0.3\hsize]{fig8b.ps}
\includegraphics[angle=-90.,width=0.3\hsize]{fig8c.ps}}
\caption{Various slit profiles of run D and S. From top to bottom:
surface density, mean LOS velocity, velocity dispersion, $h_3$, and
$h_4$. (a) left column: slit profiles parallel (thick curve) and
perpendicular (thin curve) to the major axis of the {\it primary} bar
in run D; (b) middle column: as in (a), but slits are parallel and
perpendicular to the major axis of the {\it secondary} bar in run
D. (c) right column: as in (a), but slits are parallel and
perpendicular to the major axis of the (single) bar in run S at
$t=400$.}
\label{fig:slits}
\end{figure*}

\subsection{Slit Kinematics}

Figure~\ref{fig:slits} shows various slit profiles along the
major/minor axis of the primary/secondary bar of run D at $t=405$, and
for run S. We notice that there is no central velocity dispersion
($\sigma$) drop in our simulated double-barred systems. So the
$\sigma$-drop (as found in \citealt{ems_etal_01,ems_etal_06}) is not a
requirement, and is not always associated with the formation of a
double-barred system. More likely $\sigma$-drops are just the
signature of newly-formed (therefore dynamically ``cool'') stars
\citep{ems_etal_01,ems_etal_06}. $N$-body simulations have shown that
a $\sigma$-drop can be produced in single barred galaxies
\citep{woz_etal_03, woz_cha_06}, so it is not necessarily a unique
feature of double-barred systems.

In Figure~\ref{fig:slits}(a) there is a second reversal of $h_3$ on
the major axis of the primary bar in the nuclear region
($R\la0.2$). Such a $h_3$ feature is never found for a single bar case
(run S) at any orientation. It manifests the complex asymmetric LOS
velocity distribution in the nuclear region, and is an indication of
the decoupled secondary bar. However, the reverse argument is invalid:
we found that such $h_3$ features are not visible at all orientations.

Although our secondary bar is clearly decoupled from the primary bar,
the maximum of the rotational velocity does not occur in the nuclear
region (Figure~\ref{fig:slits}), instead the rotational velocity just
rises smoothly past the region of the secondary bar (the half-length
of the secondary bar is around 0.4, see \S\ref{sec:barlength}). This
is different from \citet{ems_etal_01}, possibly indicating the
location of the maximum of the rotational velocity is not a crucial
factor for maintaining a nuclear bar.


\section{Pattern Speed Determination}

The dynamical state and evolution of barred galaxies is determined by
the pattern speed of their bars.  Knowledge of the pattern speeds of
secondary bars may constrain mechanisms of their formation and
evolution.  Not much is yet known with certainty about secondary bar
pattern speeds.  Observationally, the only direct kinematic
constraint on \omg{s}, based on the Tremaine-Weinberg
\citep{tre_wei_84} method, was obtained for NGC 2950 by CDA03, who
showed that the primary and nuclear bars cannot be rotating at the
same rate.  They further suggested that the nuclear bar of NGC 2950 is
either rotating faster or counter-rotating with
respect to the primary bar.  \citet[hereafter M06]{maciej_06}, more
emphatically, argued that NGC 2950 {\it has} to be counter-rotating
with respect to the primary.  This would raise the 
prospect that either NGC 2950 is atypical or that counter-rotating (or 
possibly librating) double bars are common.  However, this conclusion is 
based on the assumption 
that the Tremaine-Weinberg (TW) method continues to hold for nested 
bars, which CDA03 suggested may not be the case.

Our simulations provide an ideal testbed for assessing the reliability
of measurements of \omg{s}.  Here we test whether the simple version
of the TW method as used by CDA03 using 3 slits can recover \omg{s} 
accurately and check whether the signature of apparent counter-rotation
can occur without actual counter-rotation.

\subsection{The Tremaine-Weinberg method}

The TW method requires that the continuity equation is satisfied for
some kinematic tracer and that the tracer's density can be written as
$\Sigma(r,\phi-\omg{} t)$.  For slits parallel to the major axis of
the disk, if \kin{} is the luminosity-weighted mean velocity along any
such slit, and \pin{} the luminosity-weighted mean position along the
same slit, then plotting \kin{} versus \pin{} results in a straight
line with a slope of $\omg{} \sin i$.  This TW method has been used to
measure pattern speeds in large-scale bars \citep{mer_kui_95,
gersse_etal_99, deb_cor_agu_02, agu_etal_03, gersse_etal_03,
deb_wil_04}.  CDA03 showed that slits passing through the secondary
bar did not lie on the same line as those passing through only the 
primary bar, proving that $\omg{s} \neq \omg{p}$.

CDA03 argued that while \omg{p}\ can be measured from the region
outside the secondary bar, \omg{s} cannot be obtained as easily.
Their reasons for this were two-fold: (1) disentangling the
contribution to \pin{}\ and \kin\ from the primary and secondary bars
is non-trivial and (2) the secondary bar cannot be in rigid rotation
\citep{lou_ger_88, deb_she_07}, violating the assumption of the TW
method that the density can be expressed as $\Sigma(r,\phi-\omg{} t)$.
Disentangling the different contributions to integrals may be
possible: CDA03 presented two  models for doing this and M06 presented 
another.  M06 also estimated the effect of non-rigid
rotation to be less than 15\%; this estimate was however based on the 
simplifying assumption that the system is 2-D.
We therefore explore the effect of non-rigid rotation on measurements
of \omg{s} directly with the simulations.  The novelty of our approach
lies in our ability to cleanly disentangle the primary and secondary
bars: in our simulations we distinguish between disk and bulge
particles.  While the disk particles support both the secondary and
primary bars, the bulge particles almost exclusively support only the
secondary bar.  Thus if we consider only bulge particles we have a
quite clean tracer population for the nuclear bar.  It is very
unlikely that any scheme that can be devised for observational data
will ever be able to separate the secondary bar from the primary as
cleanly as we can in our simulations.

\subsection{Simulated TW measurements}

We therefore apply the TW-method to bulge particles only.  We used 11
slits covering the full region $-Y_{max} \leq Y \leq Y_{max}$, with
$Y_{max} = 0.3$ where the nuclear bar is strongest.  This corresponds
to slit-widths $\delta Y = 0.055$ or $a_{B2}/\delta Y = 7.3$ (in
comparison, the observations of CDA03 had $a_{B2}/\delta Y = 6.4$).
We adopted an inclination $i = 45\degrees$ and varied
the nuclear bar PA relative to the inclination axis, \panuc, in the range
$0\degrees \leq$ \panuc\ $\leq 90\degrees$.  We measured \pin{} and
\kin{} for each slit as in \citet[][hereafter D03]{debatt_03}:
\begin{equation}
\pin{} = \frac{1}{N_{slit}}\sum_{i \in {\rm slit}} X_i, \ \ \
\kin{} = \frac{1}{N_{slit}}\sum_{i \in {\rm slit}} V_{z,i},
\end{equation}
where $V_{z,i}$ and $X_i$ are the line-of-sight velocity and $X$
coordinate of particle $i$ and $N_{slit}$ is the number of bulge
particles in the slit.  The sums in these definitions are over bulge
particles only.

To measure \omg{s} we fit a straight line to \kin{} as a function of
\pin{} using least-squares.  As in D03, we estimate errors on the
slit integrals, \sigpin\ and \sigkin, by their radial
variation outside $|X| = 0.4$. We also experimented with a number of
other error estimates including equal errors, the difference between
positive and negative $Y$ and errors proportional to $N_{slit}$.  We
found that the most accurate measurements were obtained assuming weights
$\sigkin^{-2}$, which we adopt throughout.  In observations the main
uncertainties are in \kin\ and linear regression is dominated by
\sigkin, as here.  We denote the slope of the fitted line as $\omtw
\sin i$ in order to distinguish \omtw\ from the pattern speed,
\omg{s}, measured through the time evolution of the simulation.  We
quantify the typical errors in \omtw\ as
\begin{equation}
\sigom = \left< \left| \frac{\Delta \omg{}}{\omg{s}} \right| \right> = 
\left< \left| 1 - \frac{\omtw}{\omg{s}} \right| \right>,
\end{equation}
where $\left<\right>$ represents an average over the range $30\degrees
\leq \panuc \leq 60 \degrees$, which are favorable orientations
because they give large values of \pin{}.

\subsection{Precision of TW measurements for nuclear bars}

\begin{figure}
\includegraphics[angle=0.,width=1.\hsize]{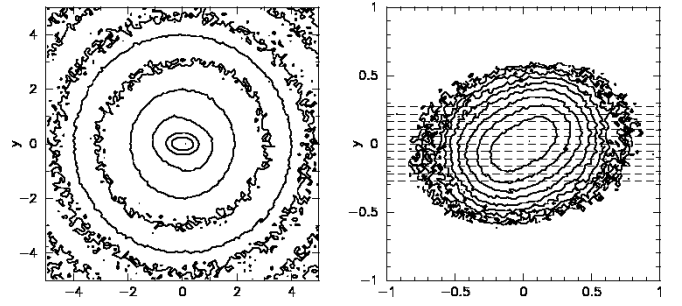}
\caption{Left: The face-on surface density (disk+bulge) in run D at
$t=20$.  Right: The bulge viewed at $i = 45\degrees$ and $\panuc =
45\degrees$ with the slits used in the TW measurement indicated by the
dashed lines.
\label{fig:sys1081_020}}
\end{figure}

\begin{figure}
\includegraphics[angle=-90.,width=1.\hsize]{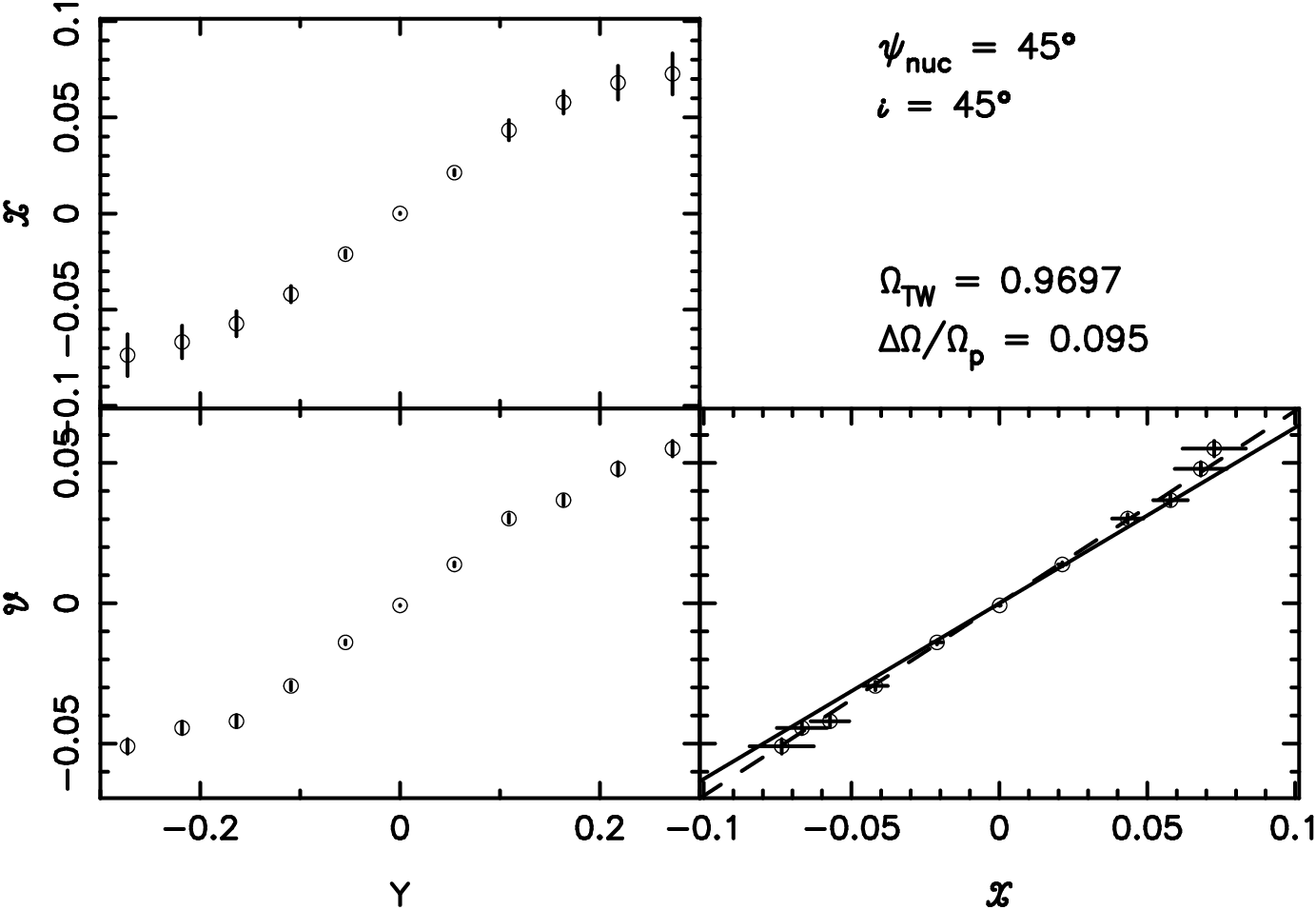}
\caption{Left: The TW integrals as a function of slit offset $Y$ for
run D at $t=20$ with $i = 45\degrees$ and $\panuc = 45\degrees$.
Right: the measurement of \omtw.  The solid line shows \omg{2}\
while the dashed line shows \omtw.
\label{fig:tw1081_020}}
\end{figure}

\begin{figure}
\includegraphics[angle=-90.,width=1.\hsize]{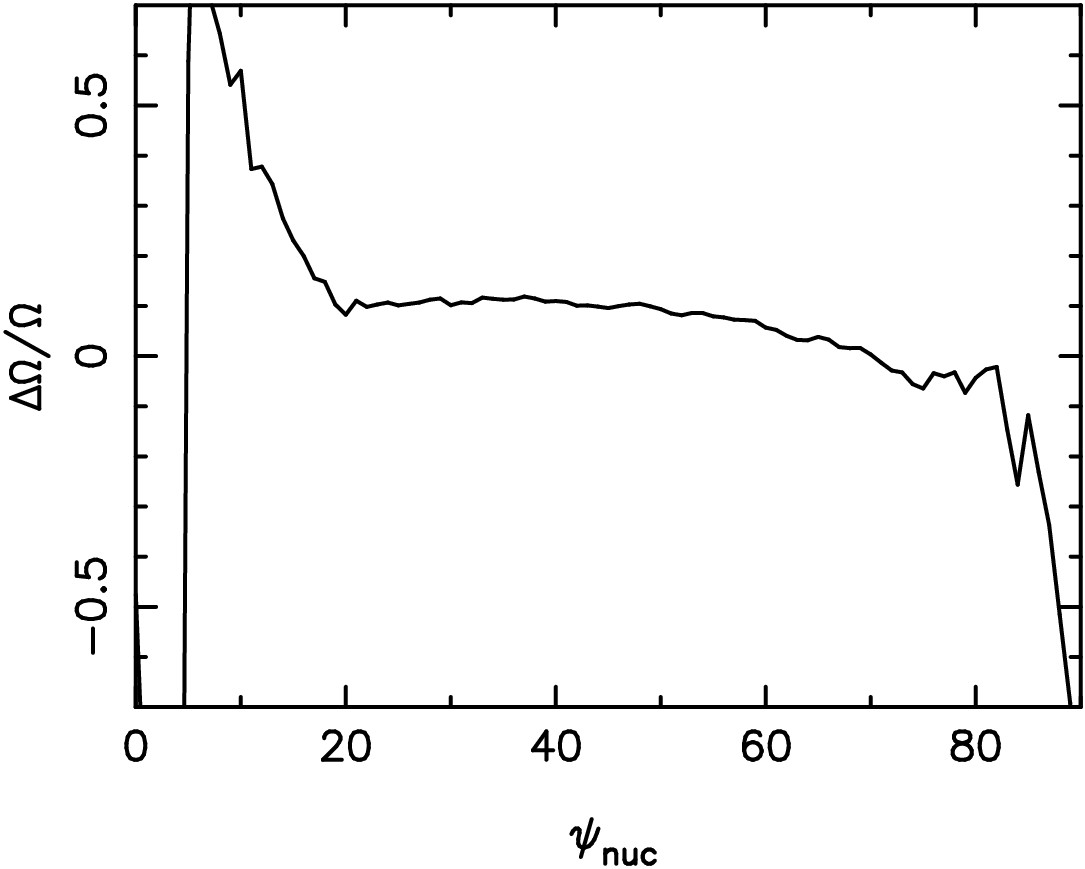}
\caption{The fractional error in \omtw\ at $t=20$ in run D as a
function of \panuc.  The average absolute error over $30\degrees \leq
\panuc \leq 60\degrees$ is $<10\%$.
\label{fig:avgtwerr1081.t20}}
\end{figure}

We start by considering the precision with which \omg{s} can be
measured in the absence of a primary bar by considering run D at
$t=20$, before the primary bar forms but after the nuclear bar has
saturated.  Figure \ref{fig:sys1081_020} presents the surface density
of the system; only a nuclear bar is present which is well traced by
the bulge particles.  The right panel shows the projected surface
density at $i = 45\degrees$ and $\panuc = 45\degrees$ with the slits
used superposed.  The value of \omg{s} measured from the time
evolution is listed in Table \ref{tab:twmeasurements}. In Figure
\ref{fig:tw1081_020} we present the TW measurement for the same 
orientation.  The measured \omtw\ is accurate to
better than 10\%, which is the typical uncertainty for single bars
\citep[D03,][]{one_dub_03}.  The integrals \pin\ and \kin\ are both
well-behaved and each pair of slits at $\pm Y$ are consistent with a
single straight line that matches the pattern speed very well.  Figure
\ref{fig:avgtwerr1081.t20} summarizes the reliability of TW
measurements for a single nuclear bar, which shows that \omg{s} can be
measured to better than $10\%$ for all reasonable orientations.  We also 
experimented with a three slit configuration consisting of the central 
$Y=0$ slit and $Y=\pm Y_{\rm offset}$ for each $Y_{\rm offset}$ and
found \sigom\ increases but is still $<14\%$.  Thus 
the TW method is well-behaved for an isolated nuclear bar.

\begin{figure}
\centerline{
\includegraphics[angle=0.,width=0.5\hsize]{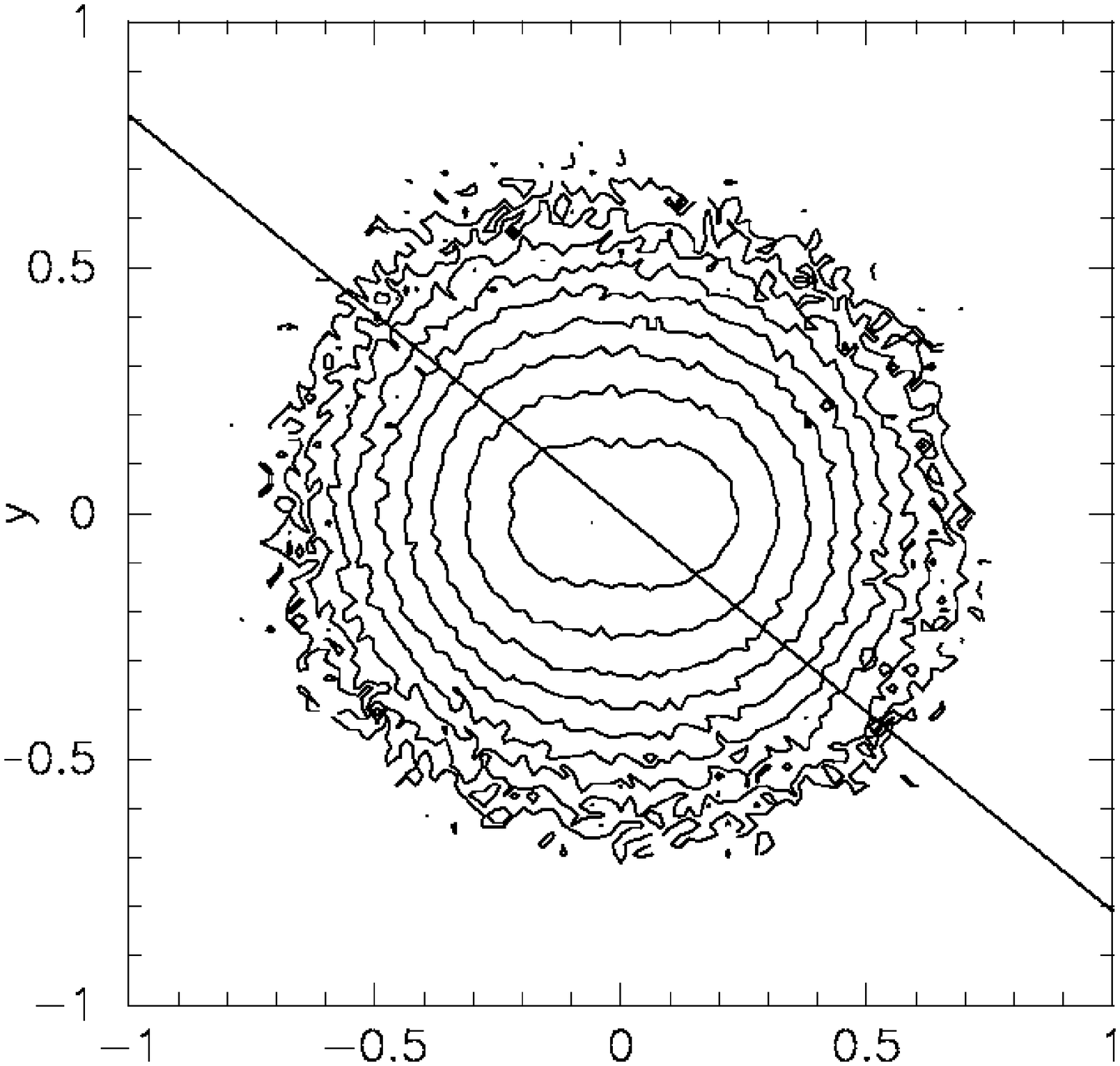}
\includegraphics[angle=0.,width=0.5\hsize]{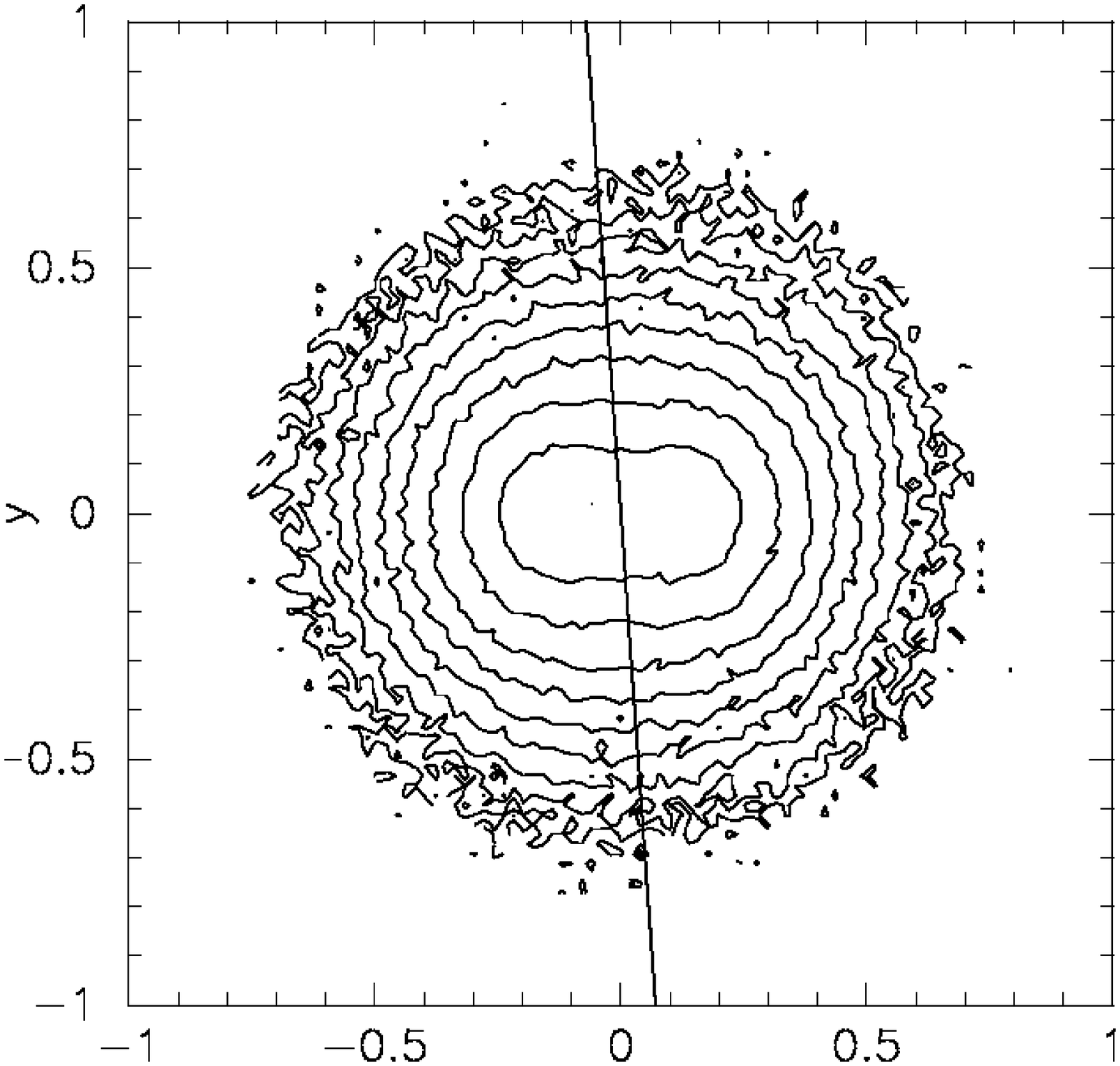}}
\centerline{
\includegraphics[angle=0.,width=0.5\hsize]{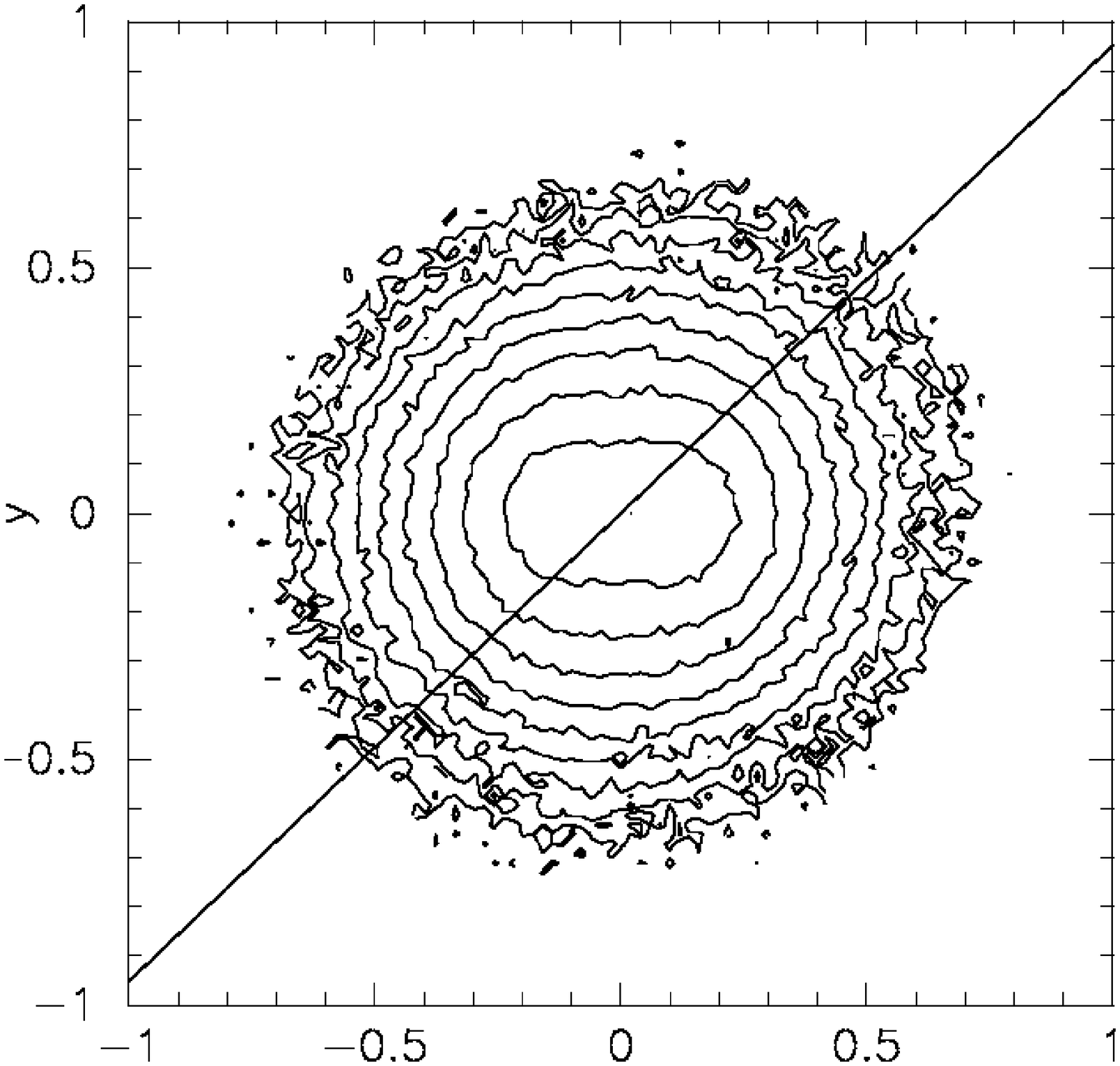}
\includegraphics[angle=0.,width=0.5\hsize]{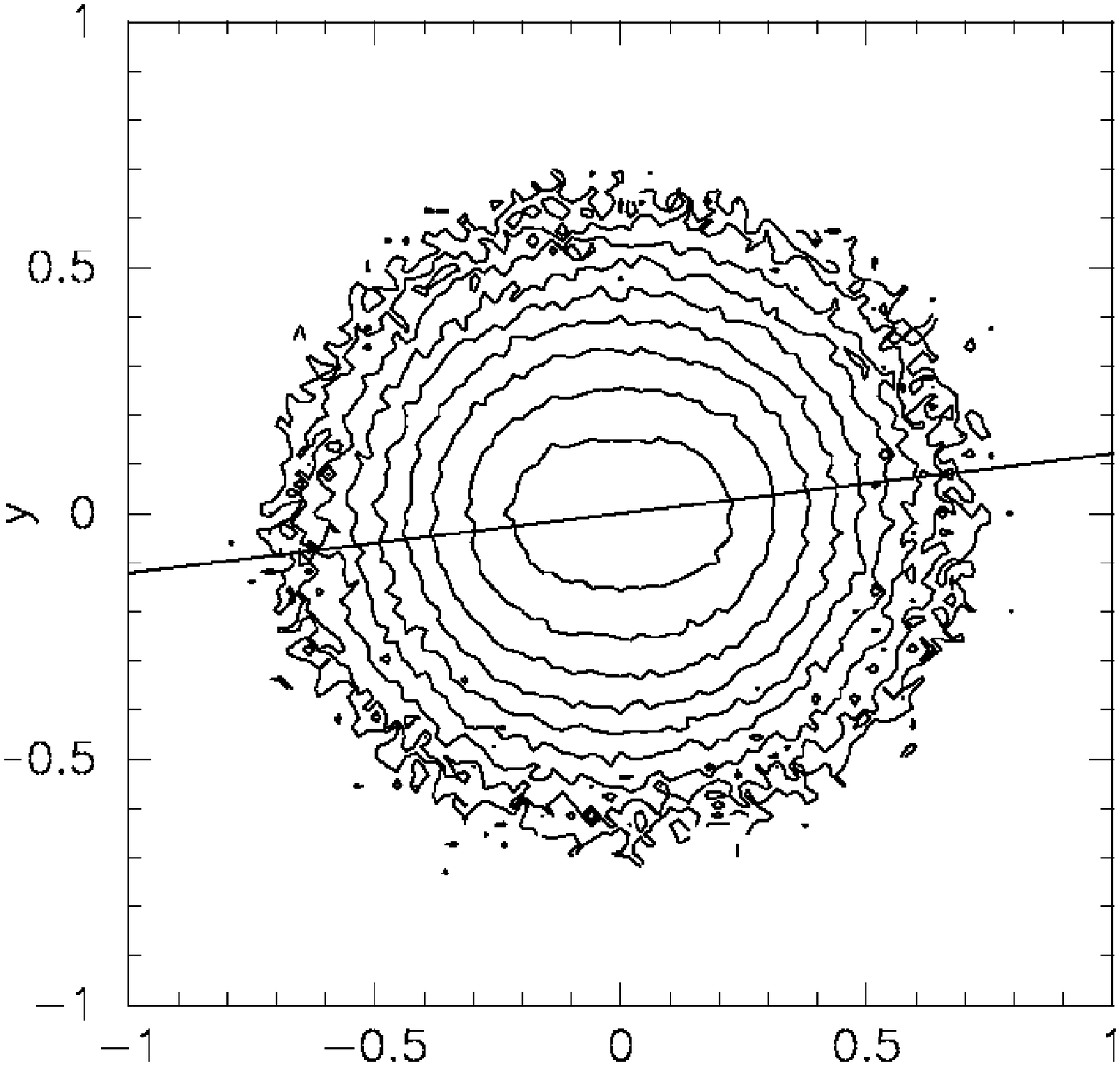}}
\caption{The face-on surface density of only bulge particles in run D.
The different snapshots are at $t=398$ (top left), $t=405$ (top
right), $t=412$ (bottom left) and $t=415$ (bottom right). The line
indicates the orientation of the primary bar at the given time.
\label{fig:s2bconts}}
\end{figure}

In Figure \ref{fig:s2bconts} we show the bulge of run D at
$t=398-415$; the nuclear bar is prominent, and has insignificant or no
elongation along the primary bar.  Thus using bulge particles only for
TW measurements will result in only the nuclear bar being included.

In Figure \ref{fig:tw4545} we present a single TW measurement at $i =
45\degrees$ and $\panuc = 45\degrees$ for each of the 4 times of
Figure \ref{fig:s2bconts}.  While the integrals themselves appear
generally well-behaved, the scatter of the points about the fitted
lines is larger than at $t=20$.  This scatter leads to a larger
\sigom\ than in the single bar case for all orientations of the
secondary bar, as shown in Figure \ref{fig:fulls2btw} and summarized
in Table \ref{tab:twmeasurements}.  Other than being large at the 
smallest values of $|\pin{}|$, \sigom\ does not correlate with 
$|\pin{}|$ or $|\kin{}|$.  When we fit lines to three slits
as before, we find that the quality of the fits varies considerably 
(Figure \ref{fig:sliterrs}).   We conclude that observationally it is 
difficult to determine the uncertainty in any measurement of \omg{s} 
based on slit data obtained by CDA03 for NGC 2950.

\begin{figure*}
\centerline{
\includegraphics[angle=-90.,width=0.5\hsize]{fig13a.ps}
\includegraphics[angle=-90.,width=0.5\hsize]{fig13b.ps}
}
\centerline{
\includegraphics[angle=-90.,width=0.5\hsize]{fig13c.ps}
\includegraphics[angle=-90.,width=0.5\hsize]{fig13d.ps}
}
\caption{TW measurements with $i=45\degrees$ and $\panuc = 45\degrees$
at $t=398$ (top-left), $t=405$ (top-right), $t=412$ (bottom-left) and
$t=415$ (bottom-right).  The solid lines show \omg{2}\ while the dashed 
lines show \omtw.
\label{fig:tw4545}}
\end{figure*}

\begin{figure}
\centerline{
\includegraphics[angle=-90.,width=\hsize]{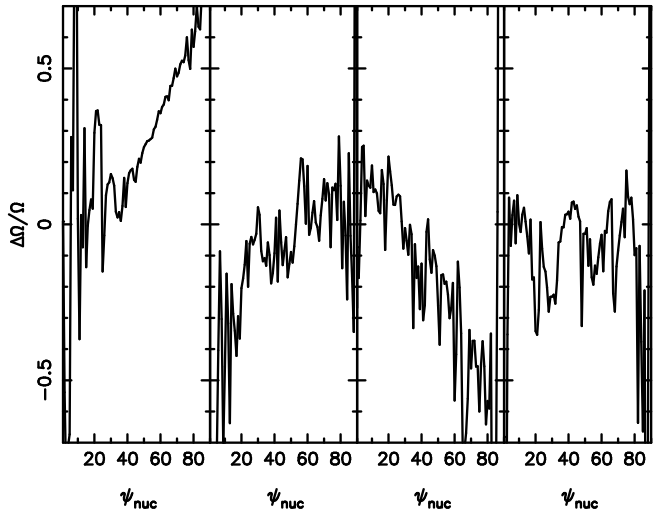}
}
\caption{The full TW analysis for only bulge particles in run D.  The
different panels are at $t=398$, $t=405$, $t=412$ and $t=415$,
respectively (from left to right).
\label{fig:fulls2btw}}
\end{figure}

\begin{figure}
\centerline{
\includegraphics[angle=-90.,width=1.\hsize]{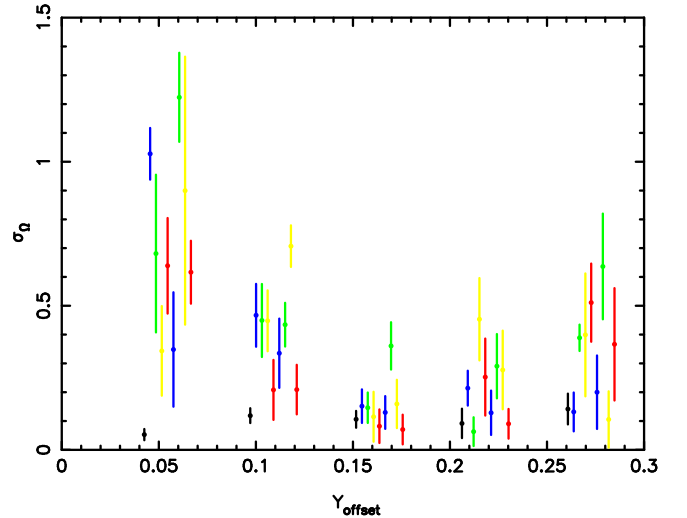}}
\caption{The precision of TW measurements of nuclear bars in 9 models
using 3 slits as described in the text.  The different points are
offset horizontally for clarity.  The colors indicate $t=20$: black,
$\panuc\simeq135\degrees$: blue ($t = 398$ and $t = 593$), $\panuc
\simeq 90\degrees$: green ($t = 405$ and $t = 599$), $\panuc \simeq
45\degrees$: yellow ($t = 412$ and $t = 605$), and $\panuc \simeq
0\degrees$: red ($t = 415$ and $t = 609$).  The error bars on the
individual points show the $1\sigma$ variations in the interval
$30\degrees \leq \panuc \leq 60\degrees$.
\label{fig:sliterrs}}
\end{figure}

\begin{table}[!ht]
\begin{centering}
\begin{tabular}{cccc}\hline
\multicolumn{1}{c}{t} &
\multicolumn{1}{c}{$\Delta \phi$} &
\multicolumn{1}{c}{\omg{s}} &
\multicolumn{1}{c}{\sigom} \\ \hline
 20 &               - & 0.89 & $0.10 \pm 0.02$  \\
398 & $ 135\degrees$  & 0.42 & $0.19 \pm 0.10$  \\
405 & $  90\degrees$  & 0.32 & $0.10 \pm 0.06$  \\
412 & $  45\degrees$  & 0.41 & $0.17 \pm 0.12$  \\
415 & $   0\degrees$  & 0.52 & $0.09 \pm 0.08$  \\ \hline
\end{tabular}
\caption{The results of TW measurements.  The column $\Delta \phi$
gives the approximate angle between the two bars.}
\label{tab:twmeasurements}
\end{centering}
\end{table}

\subsection{Interpretation}

We have demonstrated that the standard TW method on the secondary bar, 
while not wholly unreliable, is unable to recover \omg{s} without 
significant uncertainty.  Observationally this situation would be
exacerbated by the need to subtract the contribution of the primary
bar from the measured integrals, which we have not addressed \citep[but 
see][]{mei_etal_07}.

The amplitude of the $m=2$ perturbation in bulge particles varies by 
some $\pm 20\%$ about the mean amplitude at all radii.  Is the failure 
of the TW method for secondary bars consistent with the idea that 
non-rigid rotation leads to large errors?  
Evidence that this is indeed the case can be found in Figure
\ref{fig:fulls2btw}, which shows that the largest errors occur for
$\Delta \phi = 45\degrees$ and $\Delta \phi = 135\degrees$.  Figure 2
of \citet{deb_she_07} shows that the amplitude of the secondary bar,
$\amp{2} \sim -cos(2\Delta \phi)$.  Thus $d{\amp{2}}/dt$ peaks at
$\Delta \phi = \pm 45\degrees$, which is in excellent agreement with
the phases where we find the largest errors.  Moreover, the
redistribution of material being radial along the secondary bar, we
expect that the largest errors will occur when the radial motions
contribute more to the line-of-sight velocity.  While some of the error 
at all times in Figure \ref{fig:fulls2btw} is clearly due to noise, a 
significant part is also physical.  Most importantly, we find that, 
for $\Delta \phi = \pm 45\degrees$, the larger $\psi_{nuc}$ is, the 
larger is the error in \omtw.  This leads us to conclude that, as 
argued by CDA03, the perturbations to the TW method due to non-rigid 
rotation are sufficiently large as to render simple measurements of 
$\omg{2}$ noisy at best.

We have focused here on using slits to compare with the
observations of CDA03 and used only bulge particles to isolate the
nuclear bar. \citet{mei_etal_07} present an analysis using an
extension of the TW method which is able to disentangle multiple
pattern speeds provided full 2-D velocity fields.  They find, as here, 
that the pattern speed of the secondary is prone to larger uncertainties.
However, regularization with that method leads to more accurate 
measurements of \omg{s}.

\subsection{Comparison with NGC 2950}

Although we have shown that the TW method as used by CDA03 is not very 
accurate for secondary bars, it is not so grossly unreliable that we 
cannot consider the question of whether the secondary bar in NGC 2950 
is counter-rotating.  If it were, this would suggest a formation scenario 
for double-barred galaxies different from the one presented here.  
Simulations have found that counter-rotating nuclear bars are possible 
if counter-rotating material is present in the disk 
\citep{sel_mer_94, friedl_96, dav_hun_97} and this remains a viable 
model if such material is present in a sufficiently large fraction 
of galaxies.

Starting from the assumption that both bars satisfy the continuity
equation and are in rigid rotation (\ie\ $\Sigma_i(R-\omg{i}t)$ for $i
= s,p$), CDA03 showed that the TW method for the two bars combined
becomes
\begin{equation}
\label{eqn:s2btw}
(\pin{p}\omg{p} + \pin{s}\omg{s})\sin i = \kin{p} + \kin{s} \equiv \kin{}.
\end{equation}
The observed quantities are \kin{} and $\pin{} \equiv \pin{p} +
\pin{s}$ whereas the required quantities for determining the \omg{i}'s
are $\pin{i} = \int X \Sigma_i~dX$ and $\kin{i} = \int V_{los}
\Sigma_i~dX$.  Since slits can be selected to pass through the primary
but not the secondary bar, it is possible to derive \omg{p}\ assuming
that the oscillations in the primary are small (in good agreement with
our simulations).
Ignoring the effect of non-rigid rotation, CDA03 considered two
assumptions for \pin{s} in Eqn. \ref{eqn:s2btw} to solve for \omg{s}
in NGC 2950.  This gave a range of possible values of \omg{s},
including a secondary bar counter-rotating relative to the primary
bar.  Using the same data, M06 made a different attempt at isolating
the secondary bar.  Based on his analysis, M06 also argued that the
secondary bar in NGC 2950 is counter-rotating.  Since the analyses of
both CDA03 and M06 ignored the non-rigid rotation, neither of the
estimates for \omg{s} is likely to be very accurate as we showed
above.

\begin{figure}
\includegraphics[angle=-90.,width=1.\hsize]{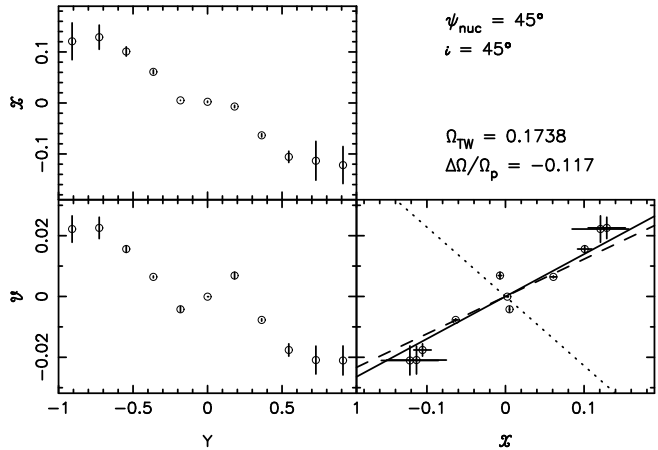}
\caption{A TW measurement for the full system of bulge+disk at $t=405$
shown in Figure \ref{fig:n2950mod}.  The solid line is for \omg{p}
while the dotted line is for $-\omg{s}$.  The best fit straight line
to the points is shown by the dashed line.  The relative weights of
bulge particles (their ``mass-to-light'' ratio) have been adjusted to
roughly reproduce $\pin{}(Y)$ in NGC 2950.  This figure is to be
compared with Figure 3 of CDA03.
\label{fig:xyv405wdisk}}
\end{figure}

Nevertheless, we do not find in our simulations cases where the
behavior of the integrals resembles that in NGC 2950.  As emphasized
by M06, the main characteristic of the $\kin{}(Y)$ profile in NGC 2950
is that it becomes steeper without changing sign in the nuclear bar
region (see Figure 3 of CDA03).  This happens despite the fact that
the two bars are on opposite sides of the minor axis (see Figure
\ref{fig:n2950mod}), causing \pin{s} to have the opposite sign of
\pin{p} and leading to $|\pin{}|$ declining more rapidly in the
secondary bar region.  But instead of \kin{}\ also being shallower in
this region, CDA03 found that $\kin{}(Y)$ steepens there.  For TW
measurements of the system in Figure \ref{fig:n2950mod}, a TW
measurement using slits passing through both bars (now with both disk
and bulge particles included) does not show a steeper $\kin{}(Y)$
profile (see Figure \ref{fig:xyv405wdisk}). We conclude that NGC 2950
may indeed have a counter-rotating primary and secondary bars. Another
possibility might be that the nuclear bar librates about the primary
bar, which deserves more investigation in future studies.

\section{Conclusions}

We have analyzed the photometrical and kinematical properties of our
high resolution models, and contrasted them when with or without a
secondary bar. This study also compared the simulated secondary bars
with observations.

In general the shape of secondary bars in our models is reasonable
compared to observed ones. The length ratio of two bars, determined by
various methods, is in the range of 0.12 to 0.19, in good agreement
with \citet{erw_spa_02, erw_spa_03}. We also found the overall edge-on
shape of boxy bulges is largely unaffected by the existence of a
secondary bar. At lower inclinations, the central density profile
along the primary bar major axis is lower than that along the minor
axis, due to the secondary bar orienting to a different direction.

The primary extends roughly to its corotation radius, and therefore fits
the definition of a fast bar (see for example
\citealt{agu_etal_03}). Although the secondary bar rotates more
rapidly than the primary, its semi-major axis is much shorter than its
corotation radius, even if we take the oscillation of the bar patterns
speeds into account. We did not find evidence of CR-ILR coupling
\citep[e.g.][]{pfe_nor_90,fri_mar_93} in our models.

We find that the central twist of kinematic axes is quite weak even if
a secondary bar is present, due to the relatively large velocity
dispersion of stars in the central region. This is consistent with the
2-D stellar kinematics of secondary bars studied in
\citet{moi_etal_04}. There are no clear $h_4$ signatures associated
with the presence of secondary bars. A $h_3$ reversal feature may
appear in the nuclear region at some favorable orientations. We do not
find a $\sigma$ (velocity dispersion) drop for our secondary bar
model. It is more likely that $\sigma$-drops are just the signature of
newly-formed stars, and it is not necessarily a unique feature of
double-barred systems.

We showed that the Tremaine-Weinberg method is not very reliable 
even when the primary bar contribution is fully excluded.  The 
way in which the measurement fails is consistent with the proposal 
of CDA03, namely that the non-rigid rotation leads to internal 
motions that violate the stationary frame assumption of the method.  
Nonetheless, we find no example in our simulations where the 
behavior of the TW integrals mimics that observed in NGC 2950.  
Thus this galaxy may indeed have counter-rotating secondary and 
primary bars.  

The general agreement between our simulations and observations of
double barred galaxies gives us confidence that the simulations are
capturing the same dynamics as in nature.  This is especially
remarkable because secondary bars are not merely scaled down versions
of primary bars, but have distinctly different kinematic properties.
In the absence of self-consistent simulations, earlier orbit-based
models could not directly confront the challenge from observations
which found such differences.  This demonstrates the advantage of
finally being able to simulate stellar double-barred galaxies, which
had been puzzling for so long.

\acknowledgements J.S. acknowledges support from a Harlan J. Smith
fellowship.  V.P.D. was supported by a Brooks Prize Fellowship at the
University of Washington and received partial support from NSF ITR
grant PHY-0205413.


\end{document}